\documentclass[nohyper,12pt,letterpaper]{JHEP3}
\usepackage{epsf}
\usepackage{epsfig}
\usepackage{cite}

\def\be{\begin{equation}}       \def\eq{\begin{equation}}
\def\ee{\end{equation}}         \def\eqe{\end{equation}}

\def\bea{\begin{eqnarray}}      \def\eqa{\begin{eqnarray}}
\def\ena{\end{eqnarray}}        \def\eea{\end{eqnarray}}
                                \def\eqae{\end{eqnarray}}

\font\cmss=cmss10

\def\a{\alpha}

\def\d{\delta}
\def\e{\epsilon}           
\def\f{\phi}               

\def\h{\eta}

\def\k{\kappa}                    
\def\l{\lambda}
\def\m{\mu}
\def\n{\nu}
  
\def\p{\pi}                
\def\r{\rho}                                     
\def\s{\sigma}                                   
\def\t{\tau}

\def\x{\xi}

\def\D{\Delta}
\def\F{\Phi}
\def\G{\Gamma}

\def\L{\Lambda}

\def\cf{{\cal F}}

\def\ch{{\cal H}}

\def\cn{{\cal N}}
\def\co{{\cal O}}

\def\cv{{\cal V}}

\def\bop#1{\setbox0=\hbox{$#1M$}\mkern1.5mu
        \vbox{\hrule height0pt depth.04\ht0
        \hbox{\vrule width.04\ht0 height.9\ht0 \kern.9\ht0
        \vrule width.04\ht0}\hrule height.04\ht0}\mkern1.5mu}
\def\pa{\partial}                              

\def\>{\rangle} 

\def\<{\langle} 
\def\Dsl{D \hskip-.6em \raise1pt\hbox{$ / $ } }

\def\leftrightarrowfill{$\mathsurround=0pt \mathord\leftarrow \mkern-6mu
       \cleaders\hbox{$\mkern-2mu \mathord- \mkern-2mu$}\hfill
       \mkern-6mu \mathord\rightarrow$}
\def\dvec#1{\vbox{\ialign{##\crcr
       \leftrightarrowfill\crcr\noalign{\kern-1pt\nointerlineskip}
       $\hfil\displaystyle{#1}\hfil$\crcr}}}          
\def\hook#1{{\vrule height#1pt width0.4pt depth0pt}}
\def\leftrighthookfill#1{$\mathsurround=0pt \mathord\hook#1
       \hrulefill\mathord\hook#1$}
\def\underhook#1{\vtop{\ialign{##\crcr                 
       $\hfil\displaystyle{#1}\hfil$\crcr
       \noalign{\kern-1pt\nointerlineskip\vskip2pt}
       \leftrighthookfill5\crcr}}}
\def\smallunderhook#1{\vtop{\ialign{##\crcr      
       $\hfil\scriptstyle{#1}\hfil$\crcr
       \noalign{\kern-1pt\nointerlineskip\vskip2pt}
       \leftrighthookfill3\crcr}}}

\def\sfrac#1#2{{\vphantom1\smash{\lower.5ex\hbox{\small$#1$}}\over
       \vphantom1\smash{\raise.4ex\hbox{\small$#2$}}}} 
\def\bfrac#1#2{{\vphantom1\smash{\lower.5ex\hbox{$#1$}}\over
       \vphantom1\smash{\raise.3ex\hbox{$#2$}}}}      
\def\afrac#1#2{{\vphantom1\smash{\lower.5ex\hbox{$#1$}}\over#2}}  
\def\on#1#2{{\buildrel{\mkern2.5mu#1\mkern-2.5mu}\over{#2}}}
\def\ddt#1{\on{\hbox{\LARGE .\kern-2pt.}}#1}             
\def\tdt#1{\on{\hbox{\LARGE .\kern-2pt.\kern-2pt.}}#1}   

\newskip\humongous \humongous=0pt plus 1000pt minus 1000pt

\newif\ifdtup

\def\to{\rightarrow}

\def\pa{\partial}

\def\apm{\alpha^{\prime}}

\def\nonu{\nonumber \\{}}
\def\half{{1 \over 2}}

\def\vp{{\vec p}}

\def\cn{${\bf C}/{\bf Z}_N\ $}
\def\zn{${\bf Z}_N\ $}

\def\IZ{\relax\ifmmode\mathchoice
{\hbox{\cmss Z\kern-.4em Z}}{\hbox{\cmss Z\kern-.4em Z}}
{\lower.9pt\hbox{\cmsss Z\kern-.4em Z}}
{\lower1.2pt\hbox{\cmsss Z\kern-.4em Z}}\else{\cmss Z\kern-.4em }\fi}
\def\IC{\relax\hbox{$\inbar\kern-.3em{\rm C}$}}
\def\IR{\relax{\rm I\kern-.18em R}}
\def\bZ{{\bf Z}}

\def\bM{{\bf M}}
\def\bC{{\bf C}}

\def\bR{{\bf R}}
\def\bT{{\bf T}}
\def\bS{{\bf S}}

\title{\Large Off-Shell Interactions for Closed-String Tachyons}

\author{{Atish Dabholkar$^{\rm b,c,d}$, Ashik Iqubal$^{\rm d}$
and Joris Raeymaekers$^{\rm a}$}\\

\centerline{$^{\rm a}$School of Physics, Korea Institute for
Advanced Study,}

\centerline{207-43, Cheongnyangni 2-Dong, Dongdaemun-Gu, Seoul
130-722, Korea}

\medskip

\centerline{$^{\rm b}$Stanford Linear Accelerator Center,}

\centerline{Stanford University, Stanford, CA 94025, USA}

\medskip

\centerline{$^{\rm c}$Institute for Theoretical Physics,
Department of Physics,}

\centerline{Stanford University, Stanford, CA 94305, USA}

\medskip

\centerline{$^{\rm d}$Department of Theoretical Physics, Tata
Institute of Fundamental Research,}

\centerline{Homi Bhabha Road, Mumbai 400005, India}

\bigskip

{\rm E-mail}:\email{atish@stanford.edu, iqubal@tifr.res.in,
joris@kias.re.kr}}

\abstract{Off-shell interactions for localized closed-string
tachyons in $\bC/\bZ_N$ superstring backgrounds are analyzed  and
a conjecture for the effective height of the tachyon potential is
elaborated. At large $N$, some of the relevant tachyons are nearly
massless and their interactions can be deduced from the S-matrix.
The cubic interactions between these tachyons and the massless
fields are computed in a closed form using orbifold CFT
techniques. The cubic interaction between nearly-massless tachyons
with different charges is shown to vanish and thus condensation of
one tachyon does not source the others. It is shown that to
leading order in $N$, the quartic contact interaction vanishes and
the massless exchanges completely account for the four point
scattering amplitude. This indicates that it is necessary to go
beyond quartic interactions or to include other fields to test the
conjecture for the height of the tachyon potential.}


\preprint{\hepth{0403238} \\ KIAS-P04017
\\SLAC-PUB-10384\\SU-ITP-04-11\\TIFR-04-04}

\begin{document}

\medskip

\pagebreak

\section{Introduction}

There are a number of physical situations, for example in
cosmology, where it is necessary to deal with unstable and
time-dependent backgrounds. It is of interest to develop
calculational tools within string theory that can describe such
backgrounds in an essentially stringy way.

A useful laboratory for studying  unstable or time-dependent
backgrounds in string theory is provided by tachyons in open
string theory. These tachyons  correspond to the instabilities of
various  unstable brane configurations and their condensation is
expected to describe the decay of these unstable branes to flat
space. The static as well as time dependent aspects of such decays
have been analyzed quite extensively.

By comparison, tachyons in closed string theory, even though more
interesting physically,  have proved to be less tractable. For
these tachyons, in most cases there is no natural candidate for a
stable minimum of the potential where the tachyon fields can
acquire an expectation value. For a closed-string tachyon with a
string-scale mass it is difficult to disentangle a well-defined
potential from other gravitational effects. In some cases, as in
the case of thermal tachyon which signifies the onset of Hagedorn
transition, the mass can be fine-tuned to be very small
\cite{Sathiapalan:1987zb, Kogan:1987jd}, but the tachyon has cubic
couplings to the dilaton and other massless scalar fields.
Consequently, it sources other massless fields which considerably
complicates the dynamics and quickly drives the system into strong
coupling or strong curvature region \cite{Atick:1988si,
Dine:2003ca}. Similarly, for the open string tachyon in the
brane-antibrane system, the mass can be tuned to zero by adjusting
the distance between the brane and the antibrane
\cite{Pesando:1999hm} to be the string scale. However, at the
endpoint of condensation, the distance between brane and antibrane
vanishes and the tachyon eventually has string scale mass and the
effective field theory breaks down.

In this paper we show that for localized tachyons in the twisted
sector of the \cn  orbifold theories, some of these difficulties
can be circumvented.  The paper is organized as follows. In
section $\S{\ref{Prediction}}$, we  motivate our computation and
elaborate arguments that lead to a conjecture for the effective
height of the tachyon potential. In section $\S{\ref{CFT}}$, we
review the orbifold CFT for a single complex twisted boson and the
four-twist correlation function that is required to describe the
scattering of four tachyons. Using factorization and symmetry
arguments we calculate all three-point correlation functions
needed for our purpose. In section $\S{\ref{Smatrix}}$, we review
superstring theory on the \cn orbifold  and using the CFT results
compute the four-tachyon scattering amplitude as well as the
gauge-invariant cubic interactions between the tachyons and the
massless fields. We show that the tachyon of interest does not
source other tachyons to quartic order and its dynamics can thus
be studied independently of others in a consistent manner. In
section $\S{\ref{Factorization}}$, we fix the normalization of the
four point amplitude by factorization and show that to order
$1/N^2$, the four tachyon scattering is completely accounted for
by the massless exchanges. We conclude in section $\S{6}$ with
some comments.

\section{A Conjecture for the Height of the Tachyon Potential}\label{Prediction}

We now pursue some of the analogies of closed-string localized
tachyons and open-string tachyons with the aim of identifying a
model where explicit computations are possible.

\subsection{Analogies with Open-String Tachyons}\label{Analogy}

There are three main simplifications which make the open-string
tachyons tractable.
\begin{itemize}
   \item The tachyons are localized on the worldvolume of an unstable
brane. It is reasonable to assume that condensation of the tachyon
corresponds to the annihilation of the brane and the system
returns to empty flat space.
    \item Conservation of energy then implies Sen's conjecture
\cite{Sen:1998sm, Sen:1999mh} that the height of the tachyon
potential should equal the tension of the brane that is
annihilated.
    \item The gravitational backreaction of D-branes can be made
arbitrarily small by making the coupling very small because the
tension of D-branes  is inversely proportional to closed string
coupling $g_c$ whereas Newton's constant is proportional to
$g_c^2$. This makes it  possible to analyze the tachyon potential
without including the backreaction of massless closed-string
modes.
\end{itemize}

The twisted-sector tachyons in \cn string backgrounds are in many
respects quite analogous.
\begin{itemize}
    \item The \cn theory has the geometry of a cone with deficit
angle $2\pi(1- 1/N)$ and the twisted-sector tachyons are localized
at the tip \cite{Dabholkar:1995ai, Lowe:1995ah}. There is
considerable evidence now that condensation of these tachyons
relaxes the cone to empty flat space and thus much like the open
string tachyons, there is a natural candidate for the endpoint of
the condensation \cite{Adams:2001sv}.
    \item There is a precise conjecture for the effective height
of the tachyon potential that is analogous to the Sen's conjecture
in the open string case \cite{Dabholkar:2001if}.
     \item At large $N$, some of the relevant tachyons
are nearly massless. Therefore, one would expect  that there is a
natural separation between the string scale and the scale at which
the dynamics of the tachyons takes place and thus higher order
stringy corrections can be controlled. The  required S-matrix
elements are completely computable using orbifold CFT techniques.
\end{itemize}

Note that for the \cn tachyons we can talk about the height of the
tachyon potential  because the background is not Lorentz
invariant. Quite generally,  two closed-string CFT backgrounds
which are both Lorentz invariant cannot be viewed as two critical
points of a scalar potential at different heights. This is because
a nonzero value of a scalar potential at a critical point with
flat geometry would generate a tadpole for the dilaton and the
string equations of motion would not be satisfied. By contrast,
the \cn backgrounds are not flat because there is a curvature
singularity at the tip of the cone. It is natural to assume that
the tachyon potential provides the energy source for the
curvature. There is not dilaton tadpole because, for a conical
geometry, the change in  the Einstein-Hilbert term in the action
precisely cancels the change in the height of the potential. Hence
the total bulk action which generates the dilaton tadpole vanishes
on the solution. There is an important boundary contribution that
is nonzero and as a result there is a net change in the classical
action. This reasoning leads to a sensible conjecture for the
height of the tachyon potential.

\subsection{A Conjecture}\label{Conjecture}

String theory on the \cn orbifold background was  first considered
in \cite{Dabholkar:1995ai, Dabholkar:1995gg, Lowe:1995ah} to model
the physics of horizons in Euclidean space. Geometrically, \cn is
a cone with deficit angle $2\pi ( 1- {1\over N})$. The tip of the
cone is a fixed point of the $\bZ_N$ orbifold symmetry and there
are tachyons in the twisted sectors  that are localized at the tip
signifying an instability of the background.

A physical interpretation of these tachyons was provided by Adams,
Polchinski, and Silverstein \cite{Adams:2001sv}. They argued that
giving expectation values to the tachyon fields would relax the
cone to flat space. The most convincing evidence for this claim
comes from the geometry seen by a D-brane probe in the sub-stringy
regime. In the probe theory, one can identify operators with the
right quantum numbers under the quantum $\hat{\bZ}_N$ symmetry of
the orbifold\footnote{The quantum $\hat{\bZ}_N$ symmetry is simply
the selection rule that twists are conserved modulo $N$. The twist
field in the k-th twisted sector has charge $k$ under this
symmetry.} that correspond to turning on tachyonic vevs. By
selectively turning on specific tachyons, the quiver theory of the
probe can be `annealed' to successively go from the $\bZ_N$
orbifold to a lower $\bZ_{M}$ orbifold with $M <N$ all the way to
flat space. The deficit angle seen by the probe in this case
changes appropriately from $2\pi(1- {1\over N})$ to $2\pi (1-
{1\over M})$.

Giving expectation value to the tachyon field in spacetime
corresponds to turning on a relevant operator on the string
worldsheet.  Thus the condensation of tachyons to different CFT
backgrounds is closely related to the renormalization group flows
on the worldsheet between different fixed points upon turning on
various relevant operators. An elegant description of the RG flows
is provided using the gauged nonlinear sigma model
\cite{Harvey:2001wm} and mirror symmetry \cite{Vafa:2001ra}. The
worldsheet dynamics also supports the expectation that the cone
will relax finally to flat space. Various aspects of localized
tachyons and related systems have been analyzed in
\cite{Dabholkar:2001gz,Michishita:2001ph,DeAlwis:2002kp,Sin:2002qs,Basu:2002jt,Sarkar:2002wb,Suyama:2002ky,Martinec:2002tz,Armoni:2003va,He:2003yw,DaCunha:2003fm,Minwalla:2003hj,Sin:2003ym,GrootNibbelink:2003zj,Lee:2003ss,Moore:2004yt,Lee:2003ar,Headrick:2003yu,David:2001vm,Gutperle:2002ki}.
For a recent review see \cite{HMT}.

These results are consistent with the assumption that in the field
space of tachyons there is a potential $V(\bT)$ where we
collectively denote all tachyons by $\bT$. The $\bZ_N$ orbifold
sits at the top of this potential, the various $\bZ_M$ orbifolds
with $ M < N$ are the other critical points of this tachyonic
potential, and flat space is at the bottom of this potential. Such
a potential can also explain why a conformal field theory exists
only for special values of deficit angles. We will be concerned
here with the static properties such as the  effective height of
the potential and not so much with dynamical details of the
process of condensation.

It may seem difficult to evaluate the change in the classical
action in going from the $\bZ_N$ orbifold to the $\bZ_{M}$
orbifold but we are helped by the fact that the orbifold is
exactly conformal. Hence the equations of motion for the dilaton
and the graviton are satisfied exactly for both backgrounds. To
calculate the change in the action, let us consider the string
effective action to leading order in $\apm$
\begin{eqnarray}%
S = {1\over 2\kappa^2} \int_M \sqrt{-g}\, e^{-2\phi}\, [R +4
(\nabla_M \phi \nabla^M \phi)
-2 \kappa^2 \delta^2(x) (\nabla_\mu \bT \nabla^\mu \bar{\bT} +
V(\bT))]{\nonumber}\\
+{1\over \kappa^2} \int_{\partial{M}} \sqrt{-g}\,e^{-2\phi}\, K\,
,\qquad \label{action}
\end{eqnarray}%
where $K$ is the extrinsic curvature, $\kappa^2 \over 8 \pi$ is
Newton's constant $G$, and $V(\bT)$ denotes the tachyon potential
localized at the defect. Here $M = 0, \ldots, 9$ run over all
spacetime directions, the cone is along the 8, 9 directions, and
$\mu = 0, \ldots, 7$ are the directions  longitudinal to the eight
dimensional defect localized at the tip of the  cone. The
extrinsic curvature term is as usual necessary to ensure that the
effective action reproduces the string equations of motion for
variations $\delta\phi$ and $\delta g$ that vanish at the
boundary.

The action is very similar to the one for a cosmic string in four
dimensions. For a cosmic string in four dimensions or equivalently
for a 7-brane in ten dimensions, the deficit angle $\delta$ in the
transverse two dimensions is given by $\delta = 8\pi G \rho \equiv
\kappa^2 \rho$ where $\rho$ is the tension of the 7-brane. We are
assuming that when the tachyon field $\bT$ has an expectation
value ${\bf T}_N$, its potential supplies an 7-brane source term
for gravity such that the total spacetime is $\bM_8 \times {\bC /
\bZ_N}$ where $\bM_8$ is the flat eight-dimensional Minkowski
space. Einstein equations then imply $R= 2 \kappa^2
\delta^2(x)V(\bT)$ and a conical curvature singularity at $x=0$.
Because of this equality, there is no source term for the dilaton
and as a result the dilaton equations are satisfied with a
constant dilaton. We see that the bulk contribution to the action
is precisely zero for the solution. The boundary has topology
$\bR^8\times {\bf S}^1$. For a cone, the circle ${\bf S}^1$ has
radius $r$ but the angular variable will go from $0$ to $2\pi
\over N$. The extrinsic curvature for the circle equals $1/r$ and
thus the contribution to the action from the boundary term equals
$ 2\pi A \over N \kappa^2$. There is an arbitrary additive
constant in the definition of the action that is determined by
demanding that flat space should have vanishing action. In any
case, we are concerned with only the differences and we conclude
that in going from \cn to flat space the total change in action
per unit area must precisely equal ${ 2\pi \over \kappa^2} (1
-{1\over N})$.

In the full string theory, we should worry about the higher order
$\apm$ corrections to the effective action. These corrections are
dependent on field redefinitions or equivalently on the
renormalization scheme of the world-sheet sigma model. However,
the total contribution of these corrections to the bulk action must
nevertheless vanish for the orbifold because we know that the
equations of motion of the dilaton are satisfied with a constant
dilaton which implies no source terms for the dilaton in the bulk.
Thus, the entire contribution to the action comes from the
boundary term even when the $\apm$ corrections are taken into
account and we can reliably calculate it in a scheme independent
way using the conical geometry of the exact solution at the
boundary.

One can convert this prediction for the change in action into a
conjecture for the height of the tachyon potential. We expect that
the tachyon potential should be identified with the source of
energy that is creating the curvature singularity. Let us see how
it works in some detail. Note that a cone has a topology of a disk
and its Euler character $\chi$ equals one. Using the Gauss-Bonnet
theorem, we then conclude $\chi = {1\over 4\pi } \int_{\bC/\bZ_N}
R + {1\over 2\pi }\int_{\bS^1} K = 1$. This implies that $R= 4\pi
(1 -{1\over N})\delta^2(x)$ and we arrive at our conjecture that
\begin{equation}\label{conjecture}
V(\bT_N) = { 2\pi \over \kappa^2} (1 -{1\over N}).
\end{equation}

We are thus led to a plausible picture rather analogous to the
open-string tachyons in which  the tachyon potential $V(\bT)$
supplies the source of energy required to create a defect and flat
space is the stable supersymmetric ground state. The landscape of
the tachyon fields in the closed string case is, however, more
intricate. There are several tachyonic modes and many critical
points corresponding to cones with different deficit angles and
thus a richer set of predictions to test.

\subsection{A Model and a Strategy for Computing Off-Shell Interactions}\label{Strategy}

The potential for open-string tachyons  been analyzed using a
number of different approaches. It has been possible to test Sen's
conjecture within open string field theory  in a number of
different formalisms
\cite{Sen:1999nx,Berkovits:2000hf,Gerasimov:2000zp,Kutasov:2000qp}
both for the bosonic and the superstring. For a recent review and
a more complete list of references see \cite{Taylor:2003gn}. Some
properties of the decay process have also been analyzed exactly in
boundary conformal field theory \cite{Sen:2002nu} and in certain
toy models exactly even nonperturbatively
\cite{McGreevy:2003kb,Klebanov:2003km,McGreevy:2003ep}.

It would be interesting to similarly develop methods within
closed-string theory to test the conjecture above for the
potential of localized tachyons. For the bosonic string, the
string field theory does not have the simple cubic form as in
Witten's open string field theory \cite{Witten:1986cc}.
Nevertheless, a well-developed formalism with non-polynomial
interactions is available \cite{Zwiebach:1993ie}. Okawa and
Zwiebach have recently applied this formalism successfully in  the
level-truncation approximation \cite{Okawa:2004rh} and have found
more than $70\%$ agreement with the conjectured answer which is
quite encouraging. We will be interested here in the localized
tachyon in the superstring.  For superstrings, there is a string
field theory formalism available only for the free theory
\cite{Berkovits:1996tn} but not yet for the interacting theory so
we need to approach the problem differently.

Corresponding to (\ref{conjecture}), there is a natural object in
the worldsheet RG flows that can be identified with the tachyon
potential \cite{Dabholkar:2001wn}. For relevant flows, however,
the relation between worldsheet quantities and spacetime physics
is somewhat indirect given the fact that  away from the conformal
point the Liouville mode of the worldsheet no longer decouples. It
is desirable  to see, to what extent, (\ref{conjecture}) can be
verified directly in spacetime.

To sidestep the use of string field theory, we work instead in the
limit of large $N$ and consider the decay process that takes $\bC/
\bZ_N$ to $\bC / \bZ_{k}$ with $k= N-j$, for some small even
integer $j = 2, 4, \ldots$. We assume that the tachyonic field
$T_k$ that connects these two critical points has a well-defined
charge $k$ under the quantum $\hat{\bZ}_N$ symmetry with its mass
given by $ m_k^2 = -{2 (N-k) \over \apm N}$. This assumption is
motivated from the worldsheet mirror description of this process
\cite{Vafa:2001ra} where the relevant operator that is turned on
has a well-defined charge under the quantum symmetry. To justify
this assumption further we will check that there are no cubic
couplings between this tachyon and other nearly massless tachyons.
Therefore, giving expectation value to this particular tachyon
does not create a tadpole for other tachyons.

Because this tachyon is nearly massless we can consider its
effective dynamics much below the string mass scale by integrating
out the massive string modes and are justified in ignoring
possible string scale corrections to the effective action. The
simplest way to model the condensation process is to imagine an
effective potential
\begin{equation}\label{effective}
{2\pi \over \kappa^2} ( 1-{1\over N}) - ({2 \over \apm}) { (N-k)
\over N} |T_k|^2 + { \lambda_k \over 4} |T_k|^4.
\end{equation}
The potential  has two extrema. At $T_k =0$ it has a maximum and
its value at the maximum is given by (\ref{conjecture}) which
supplies the energy source required at the tip of the cone \cn.
There is a minimum at $|T_k|^2 = {4 (N-k) \over \apm \lambda_k N}$
and the energy at this minimum is lowered.  We would like to
identify this minimum with the cone $\bC / \bZ_k$ which implies
the prediction
\begin{equation}\label{lambda}
{1 \over \lambda_k} ({2 \over \apm})^2 ({N-k  \over N})^2  = {2
\pi \over \kappa^2}({1\over k} - {1\over N}),
\end{equation}
so that the energy at the minimum is exactly what is needed to
create the smaller deficit angle of the cone $\bC / \bZ_k$. We are
implicitly working at large N because we have assumed that the
tachyon is nearly massless and it is meaningful to talk about its
potential ignoring the $\apm$ corrections. In the large N limit,
the final prediction for the quartic term becomes
\begin{equation}\label{lambda2}
    \l_k = {\kappa^2 \over 2 \pi} ({2 \over \apm})^2 (N-k).
\end{equation}
It would be natural to set $\apm=2$ here as in most closed string
calculations but we prefer to maintain $\apm$ throughout to keep
track of dimensions and to allow for easy comparison with other
conventions.

For the consistency for this picture it is essential that the
tachyon $T_k$ does not source any other nearly massless fields
apart from the dilaton. We have already explained  that the
dilaton tadpole vanishes because the bulk action is zero for the
cone. However, if there are tadpoles of any of the  very large
number of nearly massless tachyons in the system, it would ruin
the simple picture above. Quantum $\hat{\bZ}_N$ symmetry surely
allows terms like $T_k T_k T_{N -2k}$  because the charge needs to
be conserved only modulo $N$. If such a term is present then the
tachyon $T_{N-2k}$ will be sourced as soon as $T_k$ acquires an
expectation value and its equations of motion will also have to
satisfied. We will then be forced to take into account the cubic
and quartic interactions of all such fields. Fortunately, as we
show in section \ref{Other}, even though  the cubic couplings of
this type are allowed {\it a priori}, they actually vanish because
of $H$-charge conservation. We can thus restrict our attention
consistently to a single tachyon up to quartic order.

In what follows, we proceed with this simple ansatz.  Note that
$N-k= j$ is of order one and thus the required quartic term that
is of order one. To extract the contact quartic term, we first
need to calculate the four point tachyon scattering amplitude and
subtract from it the massless exchanges.

There is a subtlety in this procedure that is worth pointing out.
We are interested in the one-particle irreducible quartic
interaction. To obtain it from the four particle scattering
amplitude, we should subtract all one-particle reducible diagrams.
Now, in string theory, an infinite number of particles of
string-scale mass are exchanged along with the massless fields of
supergravity and it would be impractical if we have  to subtract
all such exchanges. Note however, that the mass of the tachyon of
interest is inversely proportional to $N$ and there is a clear
separation of energy scales.  We are interested in the effective
field theory at these much lower energies that are down by  a
factor of $1/N$ compared to the string scale. Therefore, massive
string exchanges are to be integrated out. For tree-level
diagrams, integrating out a massive field simply means that we
{\it keep} all one-particle reducible diagrams in which the
massive field is exchanged. This generates an effective quartic
interaction in much the same way the four-fermi interaction is
generated by integrating out the massive vector boson. For this
reason, we do not need to subtract the exchanges of massive string
modes.

The procedure would then be to compute the four-point amplitude,
subtract from it all massless or nearly-massless exchanges, and
then take the string scale to infinity and $N$ to infinity keeping
fixed the tachyon mass and the external momenta to focus on the
energy scale of interest. To put it differently, if we subtract
only the massless exchanges, we are automatically solving the
equations of motion the massive fields
\cite{Bardakci:1974gv,Bardakci:1974vs,Bardakci:1975ux,Bardakci:1978an}.
This observation explains why the full formalism of string field
theory is not needed in our case as it would be for a string scale
tachyon and we can proceed consistently within effective field
theory.

An analogous large-$N$ approximation was used by Gava, Narain, and
Sarmadi \cite{Gava:1997jt} to analyze the off-shell potential of
an open string tachyon that arises in the D2-D0 system. This
tachyon signals the instability of the system towards forming a
lower energy bound state in which the D0-brane is dissolved into
the D2-brane. These authors introduce a parameter $N$ by
considering a system of a single D2 brane with $N$ D0-branes
already dissolved in it and then introduce an additional D0-brane.
The relevant tachyon is then nearly massless when $N$ is large and
one can  consistently analyze the system in effective field theory
in much the same way as we wish to do here. For the open string
tachyon also, the mass-squared is inversely proportional to $N$
and the quartic term turns out to be of order one. The potential
then has a lower energy minimum corresponding precisely to the
lowered energy of the additional D0-brane dissolved into the D2
brane.

One would hope that a similar story works for the nearly massless
closed-string tachyons but there is no {\it a priori} way to
determine the value of the quartic term without actually computing
it. Unfortunately, our computation shows that the quartic term in
this case is not of order one but much smaller, of order $1/N^3$.
We discuss the results and  implications in some detail in section
$\S{\ref{Conclusions}}$.

It is now clear that for our purpose we require the S-matrix
element for the scattering of four tachyons and the three-point
couplings of these tachyons to massless or nearly massless fields.
We in turn need the four-point and three-point correlation
functions involving the twist fields of the bosonic and fermionic
fields. The fermionic twist fields have a free field
representation and their correlation function are straightforward
to compute. The computations involving the bosonic twist fields
are fairly involved and require the full machinery of orbifold
CFT. For this reason in the next section we focus only on the CFT
of a single complex twisted boson and determine the required
correlators using CFT techniques and factorization.

\section{Bosonic CFT on \cn}\label{CFT}

The main thrust of this section will be the computation of various
three-point functions involving the bosonic twist fields. These
correlation functions enter into the cubic interactions of the
tachyon with the untwisted massless fields carrying polarization
and momenta along the cone directions. The orbifold CFT is fairly
nontrivial and the correlations cannot be computed using free
field theory. Fortunately, it turns out that all the three-point
functions required for our purpose can be extracted from
factorization of the four-twist correlation function which is
already known in the literature. This fact considerably simplifies
life.

Our starting point will be the four-twist correlation function
which has been computed in \cite{Dixon:1987qv, Hamidi:1987vh,
Bershadsky:1987fv}.  We review some basic facts about the CFT of a
single complex boson in the twisted and the untwisted sector in
sections $\S{\ref{Untwisted}}$ and $\S{\ref{Twisted}}$ and  the
relevant aspects of the four-twist correlation function in
$\S{\ref{Fourpoint}}$. We then calculate the various three-point
functions in $\S{\ref{Threepoint}}$ up to a four-fold discrete
ambiguity using factorization and symmetry arguments. The discrete
ambiguity will be fixed later by demanding BRST invariance of the
full string vertex.

\subsection{Untwisted Sector}\label{Untwisted}

We begin with a review of the Hilbert space of the untwisted
sector a complex scalar $X$ taking values in \cn. The purpose of
this section is to keep track of factors of $N$ and to collect
some formulas  on how the states in the oscillator basis split
into primaries and descendants of the conformal algebra. This will
be important later for factorization using conformal blocks.

\subsubsection{States and Vertex Operators}

States in the untwisted sector of the CFT on \cn are constructed
by projecting the Hilbert space $\ch_{\bf C}$ of  CFT on the
complex plane onto ${\bf Z}_N$ invariant states. The \zn generator
$R$ satisfying $R^N = 1$ acts on $\ch_{\bf C}$ as a unitary
operator; $R^\dagger = R^{-1}$. From this one constructs the
orthogonal projection operator
$$ P = 1/N \sum_{k=0}^{N-1} R^k $$
satisfying $P^2 =P;\ P^\dagger = P$. The \cn Hilbert space is then
$\ch_{{\bf C/Z}_N} = P \ch_{\bf C}$. Defining complexified momenta
$$
p= {1 \over \sqrt{2}}(p_8 - i p_9)\qquad \bar p= {1 \over
  \sqrt{2}}(p_8 + i p_9),
$$
we start from momentum states $|p, \bar p \rangle$ in $\ch_{\bf
C}$ normalized as
$$ \langle p', \bar p '|p, \bar p \rangle = (2 \p )^2 \d (p - p')\d (\bar p - \bar p').$$
The states
$$ |p, \bar p \rangle_N \equiv P|p, \bar p \rangle =
1/N \sum_{k=0}^{N-1}|\h^k p, {\bar \h}^k \bar p \rangle
$$
with $\h = e^{2 \p i /N}$ form a continuous basis on \cn with
normalization
\be
 _N\langle p', \bar p '|p, \bar p \rangle_N =
(2 \p )^2 \d_N (\vp, \vp ') \label{norm} \ee
where we have defined
$$
\d_N (\vp, \vp ') \equiv 1/N \sum_{k=0}^{N-1} \d (p - \h^k p')\d
(\bar p - \bar \h ^k \bar p').
$$
The completeness relation on $\ch_{\bf C/Z_N}$  reads
$$ {\bf 1} = \int_{\bf C} {d p d \bar p \over (2 \p)^2} |p, \bar p \rangle_N
\;_N\langle p, \bar p|.$$

For an arbitrary vertex operator $\co$, we denote its projection
onto the \zn invariant subspace by $[ \co ]_N$ defined by
\be
[ \co ]_N \equiv 1/N  \sum_{k=0}^{N-1} R^k \co R^{-k}.
 \ee
The vertex operators corresponding to the states $| p, \bar p
\rangle_N$ are $[ e^{i(pX + \bar p \bar X)}]_N$. Their BPZ inner
product is related to the Hermitian inner product (\ref{norm}) on
the Hilbert space by an overall normalization constant $A$ and a
sign change on one of the momenta
\be \langle [ e^{i(p'X + \bar p' \bar X)}]'_N(\infty )[ e^{i(pX +
\bar p \bar X)}]_N(0)\rangle
 = A (2 \p)^2 \d^2_N(\vp, -\vp ')
\label{BPZ} \ee
where the prime means $ \co ' (\infty) = \lim_{z \to \infty}
z^{2h} \bar z^{2 \tilde h} \co (z) $\cite{Polchinski:1998rq}. By
construction, the overall normalization $A$ should be the same as
for the CFT on {\bf C}. We will later explicitly check this from
unitarity.

The full Hilbert space in the untwisted sector is built up by
acting on the momentum eigenstates  with creation operators $\a_{-
\{m \} },\ \bar \a_{-\{ m\} },\ \tilde \a_{-\{m\} },\ \bar{\tilde
\a}_{-\{m \} }$ and then taking the $\bZ_N$ invariant
combinations. A general state can be written as
$$
\co^{\{ \bar m \} \{ \bar {\tilde m} \} }_{\{ m \}  \{\tilde m \}
} (p, \bar p) = [\prod_i (\a_{-i})^{m_i}\prod_j
(\bar\a_{-j})^{\bar m_j}\prod_k (\tilde \a_{-k})^{m_k} \prod_l
(\bar {\tilde \a}_{-l})^{m_l} \cdot  e^{i(pX + \bar p \bar X)}]_N.
$$
The $level$ of a state is defined as the pair $(M,N)$ with $M =
\sum( m_i + \bar m_i),\
  N= \sum(\tilde m_i + \bar{\tilde m}_i)$.
Of particular interest to us are the lowest level operators which
are part of the vertex operators for massless string states:
\bea
{\rm level\ } (0,0):& \co_{00}^{00} \nonu {\rm level\ }
(1,0):& \co_{10}^{00},\co_{00}^{10}\nonu
 {\rm level\ } (0,1):& \co_{01}^{00},\co_{00}^{01}\nonu
{\rm level\ } (1,1):& \co_{11}^{00},\ \co_{00}^{11},\
\co_{10}^{01}\ , \co_{01}^{10}. \label{Xops}
\eea
Their explicit definition is
\bea \co_{00}^{00} = [e^{i(pX + \bar p \bar X)}]_N&& \nonu
\co_{10}^{00} = [\a_{-1} \cdot e^{i(pX + \bar p \bar X)}]_N,&&
\co_{01}^{00} = [\tilde \a_{-1} \cdot e^{i(pX + \bar p \bar
X)}]_N\nonu \co_{00}^{10} = [\bar \a_{-1} \cdot e^{i(pX + \bar p
\bar X)}]_N,&& \co_{00}^{01} = [\bar{\tilde \a}_{-1} \cdot e^{i(pX
+ \bar p \bar
    X)}]_N\nonu
\co_{11}^{00} = [\a_{-1}\tilde \a_{-1}\cdot e^{i(pX + \bar p \bar
X)}]_N,&& \co_{00}^{11} = [\bar \a_{-1}\bar{\tilde \a}_{-1} \cdot
e^{i(pX + \bar p \bar
    X)}]_N\nonu
\co_{10}^{01} = [\a_{-1}\bar{\tilde \a}_{-1}\cdot e^{i(pX + \bar p
\bar X)}]_N,&& \co_{01}^{10} = [\bar \a_{-1} \tilde \a_{-1} \cdot
e^{i(pX + \bar p \bar
    X)}]_N\nonumber
\eea
 For example, the vertex operator for $\co_{01}^{00}$ is given
by $$ \co_{01}^{00}(p, \bar p) = i \sqrt{2 \over \a '} [\bar \pa X
e^{i(pX + \bar p \bar X)}]_N. $$

\subsubsection{Primaries and descendants}

For computing cubic vertices, We need to write the operators
(\ref{Xops}) in terms of primaries and descendants of the Virasoro
algebra. For a primary field at level $(M,N)$ we will use the
notation $\cv^{(M,N)} (p, \bar p)$\footnote{For the levels that we
need to consider, there is only one primary field at each level so
for our purposes there is no ambiguity in this notation.}.

The state at level $(0,0)$ is obviously a primary,
\be \cv^{(0,0)} = \co_{00}^{00}=[ e^{i(pX + \bar p \bar X)}]_N
\label{levelzero} \ee At level $(1,0)$, there is a primary
$\cv^{(1,0)}$ and a descendant $L_{-1} \cdot \cv_{(0,0)}$:
\bea
\cv^{(1,0)} &=&{ 1\over \sqrt{2p \bar p}}\left(\bar p
\co_{10}^{00} - p \co_{00}^{10} \right) \nonu L_{-1} \cdot
\cv^{(0,0)} &=& \sqrt{\a' \over 2} \left(\bar p \co_{10}^{00} + p
\co_{00}^{10} \right).\label{level10}
\eea
where primary fields are  delta-function normalized under the BPZ
inner product as in (\ref{BPZ}). Similarly, at level $(0,1)$ we
have:
\bea \cv^{(0,1)}  &=&{ 1\over \sqrt{2p \bar p}}\left(p
\co_{01}^{00}  - \bar p \co_{00}^{01} \right) \nonu \tilde L_{-1}
\cdot \cv^{(0,0)} &=& \sqrt{\a' \over 2} \left(p \co_{01}^{00}  +
\bar p \co_{00}^{01} \right).\label{level01}
\eea
At level $(1,1)$, we find a primary and three descendants:
\bea \cv^{(1,1)} &=&{ 1\over 2p \bar p}\left(p^2 \co_{11}^{00}  -
p\bar p \co_{10}^{01} -p \bar p  \co_{01}^{10}+ \bar p^2
\co_{00}^{11}\right)\nonu \tilde L_{-1} L_{-1}\cdot \cv^{(0,0)}
&=& {\a' \over 2}\left(p^2 \co_{11}^{00}  + p\bar p \co_{10}^{01}
+ p \bar p  \co_{01}^{10}+ \bar p^2 \co_{00}^{11}\right)\nonu
\tilde L_{-1}\cdot \cv^{(1,0)} &=& \sqrt{\a ' \over 2 p \bar
p}\left(p^2 \co_{11}^{00}  + p\bar p \co_{10}^{01} -p \bar p
\co_{01}^{10}- \bar p^2 \co_{00}^{11}\right)\nonu L_{-1}\cdot
\cv^{(0,1)} &=& \sqrt{\a ' \over 2 p \bar p}\left(p^2
\co_{11}^{00}  - p\bar p \co_{10}^{01} +p \bar p  \co_{01}^{10}-
\bar p^2 \co_{00}^{11}\right)\label{level11}
\eea

\subsection{Twisted Sector}\label{Twisted}

Twisted sector states are created by the insertion of $twist\
fields$. The bosonic twist field that creates a state in the
$k$-th twisted sector is denoted by $\s_k$. The $\s_k$ are primary
fields of weight $h= \tilde h = \half k/N(1-k/N)$. Their OPE's are
given by \bea \pa X (z)  \s_k (0) &\sim& z^{-1 + k/N} \t_k (0)+
\ldots\nonu \pa \bar X (z) \s_k(0) &\sim& z^{-k/N} \t_k' (0)
+\ldots \eea where $ \t_k,\ \t_k' $ are $excited\ twist\ fields$.
In the presence of a twist field, the mode numbers of $\pa X,\ \pa
\bar X$ get shifted: \bea \pa X &=& -i \sqrt{\a ' \over 2} \sum_m
\a_{m-k/N} z^{-m-1+ k/N}\nonu \pa \bar X &=&  -i \sqrt{\a ' \over
2} \sum_m \bar \a_{m+k/N} z^{-m-1- k/N}. \eea The state
$|\s_k\rangle \equiv \s_k(0)|0\rangle$ is annihilated by all
positive frequency modes. The commutation relations are
$$
[\a_{n-k/N}, \bar \a_{m+k/N}] = (m + k/N) \d_{m,-n}.
$$
We take the BPZ inner product between twist fields to be
normalized as \be \langle \s_{N-k}'(\infty) \s_k(0)\rangle = A
\label{sigmanorm} \ee
where $A$ is the same constant that appears in (\ref{BPZ}). This
is just a convenient choice -- a different normalization of the
twist fields can be absorbed in the proportionality constant
multiplying the vertex operators for strings in the twisted
sectors. The latter is ultimately determined by unitarity as we
discuss in $\S{\ref{Smatrix}}$.

\subsection{Four-Twist Correlation Function}\label{Fourpoint}

The four-twist amplitude  is given by
\cite{Dixon:1987qv,Hamidi:1987vh,Bershadsky:1987fv}
\be
 Z_4 (z, \bar z) \equiv \langle \s_{N-k}'(\infty) \s_k(1)
\s_{N-k}(z, \bar z) \s_k(0) \rangle = A B |z (1-z)|^{-2{k\over
N}(1 -{k\over N})} I(z, \bar z)^{-1} \label{Z}
\ee
where
$$
I(z, \bar z) = F(z) \bar F ( 1 - \bar z) + \bar F (\bar z) F(1 -
z),
$$
$$
F(z) = \;_2F_1(k/N, 1 - k/N,1,z).
$$
The overall normalization  $A$ is expressible as a functional
determinant and is common to all amplitudes of the $X,\ \bar X$
CFT on the sphere, while $B$ is a numerical factor which we will
relate, through factorization, to the  normalization of the
two-twist correlator. We will frequently need the asymptotics for
the hypergeometric function $F$:
\bea F(z) &\sim& 1,\qquad   z \to 0, \nonu F(1-z) &\sim& - {1\over
\pi} \sin {\pi k \over N}  \ln {z \over \delta},\qquad z\to 0,
\nonu F(z) &\sim& e^{ \pi i k/N} {\Gamma (1 - 2k/N) \over \Gamma^2
(1 - k/N)} z^{-k/N} - e^{-\pi i k/N} {\Gamma (2k/N-1) \over
\Gamma^2 (k/N)} z^{-(1-k/N)},\ z \to \infty,
\nonu F(1-z) &\sim&
{\Gamma (1 - 2k/N) \over \Gamma^2 (1 - k/N)} z^{-k/N} + {\Gamma
(2k/N - 1) \over \Gamma^2 (k/N)} z^{-(1-k/N)},\ z \to \infty,
\label{asymptotics} \eea
where $\d$ is defined by
\be \ln \d = 2 \psi (1) - \psi(1 - k/N) - \psi(k/N) \ee
with $\psi(z) \equiv {d \over d z} \ln \G(z)$. For  $1-k/N$ small,
$\ln \d$ behaves as
$$
\ln \d = {N \over N-k} + \co ({N-k \over N})^2.
$$

\subsection{Three-point Correlation Functions}\label{Threepoint}

We are now ready to calculate various three-point functions of two
twist fields and an untwisted field by factorizing the four-twist
amplitude\footnote{Some of the correlation functions have been
computed independently in \cite{Okawa:2004rh}.}. In a general CFT,
the four-point amplitude can be expanded as
\be Z_4 (z, \bar z) = \sum_p ( C_{-+}^p )^2 \cf(p|z) \bar
\cf(p|\bar z)
 \label{confblock}
\ee
where the sum runs over primary fields, $C_{-+}^p$ are
coefficients in the $\s_{N-k}\s_k$ OPE, and $\cf(p|z),\ \bar
\cf(p|\bar z)$ are the conformal blocks. The conformal blocks in
turn can be expanded for small $z$ as
\bea \cf(p|z) &=& z^{h_p - k/N(1-k/N)} ( 1 + \half h_p z + \co(
z^2))\nonu \bar \cf(p|\bar z) &=& \bar z^{h_p - k/N(1-k/N)} ( 1 +
\half \tilde h_p \bar z + \co( \bar z^2)). \eea
For  a discussion and derivation of this formula see for example
\cite{DiFrancesco:1997nk}.

In our case, the sum over primaries in (\ref{confblock}) is in
fact a discrete sum over primaries at different levels as well as
an integral over `momenta' $(p,\bar p)$. Therefore, to order
$|z|^2$ we have
\bea Z_4 (z, \bar z) &=& \int_{\bf C} {dp d\bar p \over (2 \p)^2
A} |z|^{\a ' p \bar p - 2 k/N(1-k/N)} \Big[ ( C_{-+}^{(0,0)}
(p,\bar p) )^2 \nonu &+& z \left( {\a '\over 4} p \bar p  (
C_{-+}^{(0,0)} (p,\bar p) )^2 + ( C_{-+}^{(1,0)} (p,\bar p)
)^2\right)\nonu &+& \bar z \left( {\a '\over 4} p \bar p  (
C_{-+}^{(0,0)} (p,\bar p) )^2 +
 ( C_{-+}^{(0,1)} (p,\bar p) )^2\right)\nonu
& +& |z|^2 \Big( ({\a' p \bar p \over 4})^2 ( C_{-+}^{(0,0)}
(p,\bar p) )^2  + ( C_{-+}^{(1,1)} (p,\bar p) )^2\nonu & +& {\a' p
\bar p \over 4} ( ( C_{-+}^{(1,0)} (p,\bar p) )^2+( C_{-+}^{(0,1)}
(p,\bar p) )^2)\Big) + \ldots\Big] \label{Z4exp1} \eea
The coefficients $C_{-+}^{(M,N)}$ are the three-point functions
$$
C_{-+}^{(M,N)}(p, \bar p) =  \langle \s_{N-k}'(\infty) \s_{k}(1)
\cv^{(M,N)} (p, \bar p)(0) \rangle
$$
that we are interested in.

Factorization implies that the above expression for $Z_4$ should
equal the expansion of (\ref{Z}) for small $z$ which is given by
\bea Z_4 (z, \bar z) &=& { A B  \p  \over 4 \sin({\p k \over
N}) } |z|^{-2k/N(1-k/N)} \left( - {1 \over \log {|z| \over \d}} +
{a (z + \bar z)\over (\log {|z| \over \d})^2} - {2 a^2 |z|^2\over
(\log {|z| \over \d})^3}+\ldots \right)\nonu &=& {\a ' A B \over 2
\sin \p k/N} \int_{\bf C} {dp d \bar p } |z|^{\a ' p \bar p - 2
k/N (1 - k/N)} \d^{- \a ' p \bar p}\nonu &&\left[ 1+ a \a'  p \bar
p (z + \bar z) + (a \a' p \bar p)^2 |z|^2 + \ldots\right]
\label{Z4exp2} \eea
where
\be a = \half \left((k/N)^2 + (1- k/N)^2\right). \ee
In the second line of (\ref{Z4exp2}), we have used the identity
$$
\left( \log {|z| \over \d} \right)^{-(n+1)} = { (-\a')^{n+1} \over
2 \p n!} \int_{\bf C} {dp d\bar p } (p \bar p)^n
 \left( {|z| \over \d} \right)^{\a' p \bar p}.
$$

Comparing \ref{Z4exp1} and \ref{Z4exp2}, we can now read off
various operator product coefficients. For the first coefficient
we find
$$ ( C_{-+}^{(0,0)} (p,\bar p) )^2 = { \p^2 \a'A^2 B
\over \sin \p k/N} \d^{- \a ' p \bar p}. $$
Using the fact that
$$ C_{-+}^{(0,0)}(0, \bar 0) = A $$
as in (\ref{levelzero}) and (\ref{sigmanorm}) we can determine the
numerical constant $B$ that appears in \ref{Z},
\begin{equation}\label{B}
    B = { \sin {\p k \over N} \over  \p^2 \a '}.
\end{equation}
Further comparing (\ref{Z4exp1}) and (\ref{Z4exp2}) determines the
higher operator product coefficients up to signs:
\bea C_{-+}^{(0,0)} (p,\bar p)
&=& A \d^{- \a ' p \bar p / 2}\nonu C_{-+}^{(1,0)} (p,\bar p) = -
C_{-+}^{(0,1)} (p,\bar p) &=& \e_1 A {\sqrt{\a ' p \bar p}\over 2}
(1 - 2 k/N) \d^{- \a ' p \bar p / 2}\nonu C_{-+}^{(1,1)} (p,\bar
p) &=& \e_2 A {\a' p \bar p \over 4} (1 - 4 k/N (1 - k/N))
 \d^{- \a ' p \bar p / 2}.
\eea
where $\e_{1,2} = \pm 1$  are the sign ambiguities that arise from
taking square roots. In the second line, we have taken the
opposite sign for the coefficients as required by worldsheet
parity, which takes $\cv_{(1,0)}\to\cv_{(0,1)}$ and $\s_k \to
\s_{N-k}$. The three-point functions involving descendants are
easily calculated from the ones involving primaries using the fact
that $L_{-1}$ and $\tilde L_{-1}$ act on vertex operators as $\pa$
and $\bar \pa$ respectively.

These results are easily transformed to the $\a$-oscillator basis.
The operators we need were denoted by $\co_{m\tilde m}^{\bar m
\bar{\tilde m}}$ in $\S{\ref{Untwisted}}$.  Using the formulas
(\ref{levelzero}-\ref{level11}) to transform to the
$\a$-oscillator basis one finds the required three-point
functions. Using the notation
$$
D_{m\tilde m}^{\bar m \bar {\tilde m}} (p, \bar p, k) \equiv
\langle \s_{N-k}'(\infty) \s_{k}(1) \co_{m\tilde m}^{\bar m \bar
{\tilde m}} (p, \bar p)(0) \rangle
$$
we find
\bea D_{00}^{00}&=& A \d^{- \a ' p \bar p / 2}\nonu D_{10}^{00}
&=&  \sqrt{\a' \over 2} A \bar p \d^{- \a ' p \bar p / 2} \left(
\half ( 1 + \e_1) - \e_1 k/N \right)\nonu D_{00}^{10} &=&
\sqrt{\a' \over 2} A p \d^{- \a ' p \bar p / 2} \left( \half( 1 -
\e_1) + \e_1 k/N \right)\nonu D_{01}^{00} &=&  \sqrt{\a' \over 2}
A \bar p \d^{- \a ' p \bar p / 2} \left( \half( 1 - \e_1) + \e_1
k/N \right)\nonu D_{00}^{01} &=&  \sqrt{\a' \over 2} A p \d^{- \a
' p \bar p / 2} \left( \half( 1 + \e_1) - \e_1 k/N \right)\nonu
D_{11}^{00} &=& {\a ' \over 8} A \bar p^2 \d^{- \a ' p \bar p / 2}
\left( \e_2 ( 1 - 4 k/N(1-k/N)) + 1\right)\nonu D_{00}^{11} &=&
{\a ' \over 8} A p^2 \d^{- \a ' p \bar p / 2}
 \left( \e_2 ( 1 - 4 k/N(1-k/N)) + 1\right)\nonu
D_{10}^{01} &=& {\a ' \over 8} A p\bar p \d^{- \a ' p \bar p / 2}
 \left( - \e_2 ( 1 - 4 k/N(1-k/N)) + 1 + 2 \e_1 (1 - 2 k/N)\right)\nonu
D_{01}^{10} &=& {\a ' \over 8} A p\bar p \d^{- \a ' p \bar p / 2}
 \left( - \e_2 ( 1 - 4 k/N(1-k/N)) + 1 - 2 \e_1 (1 - 2 k/N)\right)\label{ccoeffs}
\eea
We later argue that the correct choice for our purpose is $\e_1=
\e_2 =-1$ which makes the total cubic string vertex
BRST-invariant.

Note that this  method to calculate the three-point function from
factorization works only as long as there is only one primary
field at each level. At level $(2,0)$, for example, one finds that
there are two primaries and their correlators cannot be determined
from factorization only. Computation of these higher correlators
is substantially more difficult but fortunately we do not require
them here. We need to subtract only massless exchanges and the
higher level primary fields do not enter into the massless vertex
operators. Equipped with the four and three point correlation
functions we are thus ready to discuss the tachyon interactions.

\section{Strings on \cn}\label{Smatrix}

We now consider type IIA/B strings on $\bM^8 \times {\bf C / Z}_N$
with $\bM^8$ representing $7+1$ dimensional Minkowski space. After
reviewing the conventions and the vertex operators, we write the
four point scattering amplitude. We then write the gauge invariant
cubic interaction between the tachyon and the massless fields and
finally determine the one-particle irreducible effective quartic
interaction.

\subsection{Conventions}

We use the coordinates $X^M = (X^\m,X,\bar X)$ where $\m
=0,\ldots,7$ and $X = {1 \over \sqrt{2}} (X^8 + i X^9)$.
Similarly, the world-sheet fermions are $\psi^M = (\psi^\m, \psi,
\bar \psi)$ with $\psi =  {1 \over \sqrt{2}} (\psi^8 + i \psi^9)$.
Our metric signature is $(-++ \ldots +)$. The orbifold \cn
represents a cone with opening angle $2\p /N$. The ${\bf Z}_N$
generator is given by
$$ R = \exp (2 \p i {N + 1 \over N} J_{89}). $$
where $J_{89}$ generates rotations in the $X^8,\ X^9$ plane.
We take $N$ to be odd so that $R^N=1$ on spacetime fermions and
the  bulk tachyon is projected out by GSO projection
\cite{Adams:2001sv}.

In the untwisted sector, one has to project onto
${\bf Z}_N$ invariant states, for which we use the notation
$[\ldots ]_N$:
\be
[ \co ]_N \equiv 1/N  \sum_{k=0}^{N-1} R^k \co (R^{-1})^k.
\label{symmop}
\ee
The vertex operators in the sector twisted by $R^k$ will contain
bosonic twist fields $\s_k$ and fermionic ones $s_k$. Their role
is to create branch cuts in the OPEs with $X,\ \bar X$ and $\psi,\
\bar \psi$ respectively. In the presence of a twist field, mode
numbers get shifted by an amount $k/N$.

The bosonic twisted sector has been discussed in detail in the
previous section. The fermionic twist fields are denoted by $s_k$.
Their OPEs are
\bea \psi (z) s_k (0) &\sim& z^{k/N} t_k + \ldots\nonu \bar
\psi(z) s_k (0) &\sim & z^{-k/N} t_k' + \ldots \eea The mode
expansions in the NS sector are \bea \psi (z) &=& \sum_{r \in \bZ
+ \half} \psi_{r-k/N} z^{-r-\half +
  k/N}\nonu
\bar \psi (z) &=& \sum_{r \in \bZ + \half}\bar \psi_{r+k/N}
z^{-r-\half -
  k/N}
\eea
with commutation relations
$$
\{ \psi_{r-k/N}, \bar \psi_{s+ k/N} \} = \d_{r,-s}.
$$
These twist fields have a free field representation in terms of
bosonized fields $H,\ \tilde H$. The latter are defined by \bea
\psi = e^{iH} ,&\qquad& \bar \psi =   e^{-iH}\nonu \tilde \psi =
e^{i\tilde H},&\qquad& \bar{\tilde \psi} =   e^{-i\tilde
  H}
\eea The twist fields are then represented by \be s_k = e^{i
{k\over N} H}, \qquad \tilde s_k = e^{-i {k\over N}\tilde H} \ee
In this representation, the only computationally nontrivial CFT
correlators  are the ones involving the bosonic twist fields
$\s_k$ that were calculated in the previous section. General
amplitudes are restricted by {\em quantum symmetry} $\hat{Z}_N$
and {\em charge conservation} for the $H, \tilde H$ fields.

\subsection{Tachyon Spectrum and Vertex Operators}

Let us  review the GSO projection and the tachyonic spectrum
\cite{Dabholkar:1995ai, Adams:2001sv}. The vertex operator for the
$k$-th twisted sector ground state, in the $-1$ picture, is
$$
\s_k e^{i {k \over N} (H - \tilde H)}
 e^{-\f - \tilde \f}
$$
For $k$ odd, this ground state is not projected out by the GSO
projection and the lowest lying mode is a tachyon with mass
$ m^2  = - 2/\a '(1 - {k \over N})$. Its vertex operator is
\be T_k^{(-1,-1)}
(z, \bar z, p) = g'_c e^{i p \cdot X} c \tilde c \s_k e^{i {k
\over N} (H - \tilde H)} e^{- \f - \tilde \f} (z, \bar z) \qquad k
\ odd \label{Tk} \ee
The normalization constant
$g_c'$  will be determined in terms of the closed string coupling
$g_c$ from factorization and the requirement that the tachyon
vertex operator
represents a canonically normalized field in $7+1$ dimensions.

For $k$ even, the ground state is projected out
by GSO projection and the tachyon is an excited state $\bar \psi_{-
\half + k/N} \tilde {\psi}_{-
\half + k/N} |0, p\rangle$
with vertex operator
$$
T_k^{(-1,-1)} (z, \bar z, p) = g'_c e^{i p \cdot X}  c \tilde c
\s_k  e^{i (k/N -1) (H - \tilde H)}
e^{- \f - \tilde \f} (z, \bar z) \qquad k \ even.
$$
Its mass-shell condition is ${-\a ' p^2 \over 4} = - \half {k
\over N}$.  This vertex operator is the complex conjugate of
$T_{N-k}$ and we denote it by $\bar T_{N-k}$. Henceforth, we take
$k$ to be odd and describe all tachyons by the vertex operators
$T_k,\ \bar T_k$. The most marginal tachyon has $k=N-2$ and $m^2 =
- {4 \over N \a'}$.

We will also need the tachyon vertex
operators in the $0$ picture
$$ T_k^{(0,0)}
(z, \bar z, p) = - {\a' \over 2} g'_c e^{i p \cdot X} p \cdot \psi\ p
\cdot
\tilde \psi  c \tilde c\s_k e^{i {k\over N}(H - \tilde H)}
 (z, \bar z)+ \ldots
$$
The omission stands for terms that arise from the part of the
picture-changing operator involving the \cn fields. These terms
have different $(H,\tilde H)$ charges from the one displayed. In
the amplitudes we  consider it is possible to choose pictures such
that the omitted terms do not contribute because of $H$-charge
conservation.

\subsection{Scattering Amplitude for Four Tachyons}

We are now ready to write down the four-tachyon scattering
amplitude. Using (\ref{Z}) it is given by
\bea V_4 &=& \int d^2z
\langle \bar T_k^{(-1,-1)} (z_\infty, \bar z_\infty,
p_1)T_k^{(0,0)}
 (1, 1, p_2)\bar T_k^{(-1,-1)} (z, \bar z, p_3)T_k^{(0,0)}
 (0, 0, p_4) \rangle \nonumber\\
&=& {{g'_c}^4  C  \over \pi^2  \a'}\sin({\pi k \over N})\D
 (\a' p_1 \cdot p_3/2)^2
\int d^2 z | 1- z|^{-{\a 't \over 2} -2} |z|^{-{\a' s \over 2}
 - 2} I(z, \bar z)^{-1},\nonumber\\
\label{4point}
\eea
where $C$ represents the product of $A$ with similar functional
determinants from the $X^\m$, the fermions and the ghosts.
\FIGURE{\epsfig{file=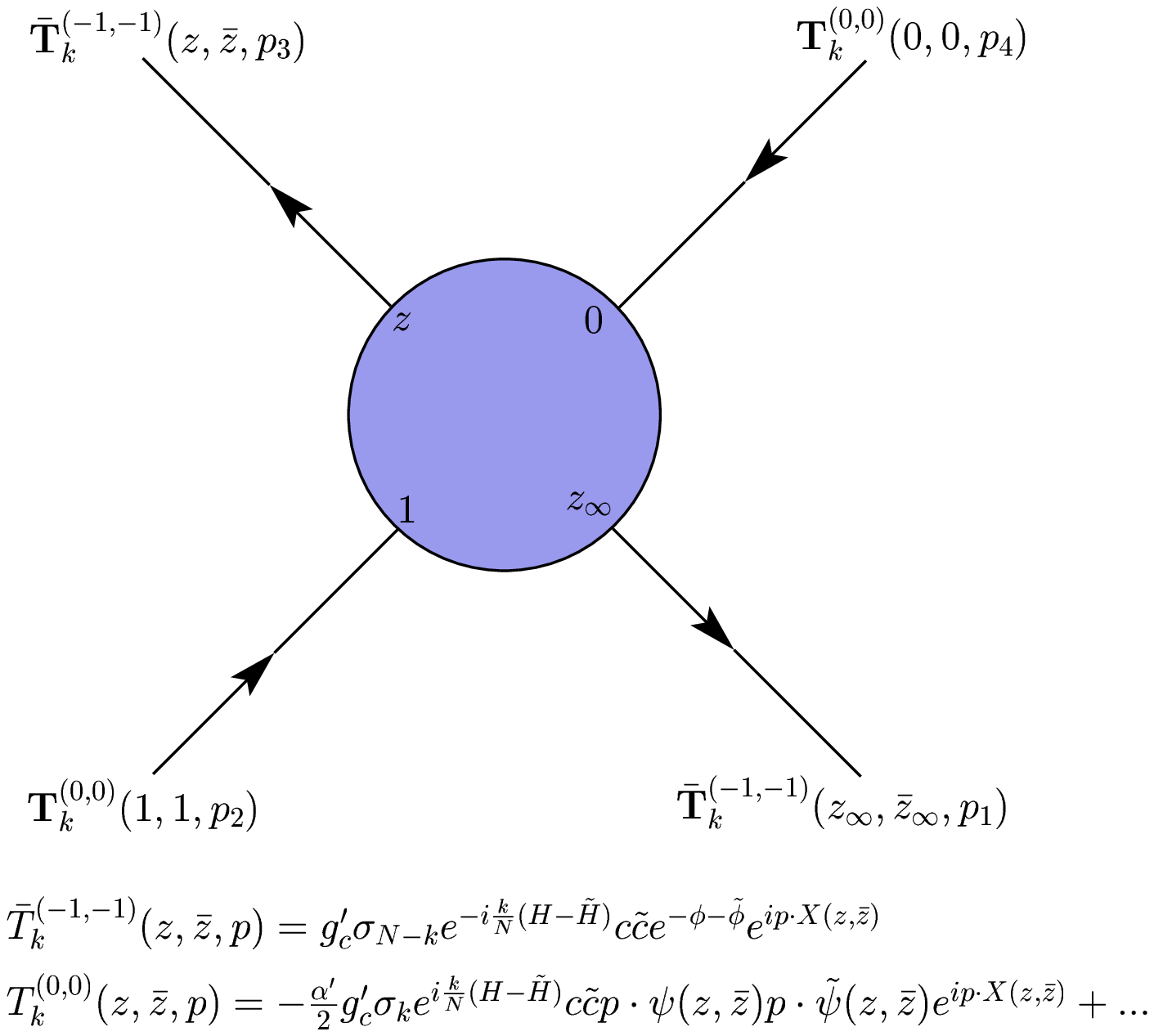,width=10cm}\caption{Four-tachyon
scattering amplitude.}}
The assignment of pictures and worldsheet positions of the vertex
operators are shown in figure 1. The Mandelstam variables $s,\ t,\
u$ are defined as
$$
s = - (p_1 + p_2)^2,\ t = - (p_2 + p_3)^2,\ u =
-(p_1 + p_3)^2
$$
and  the symbol $\D$ to denotes the 8-dimensional
momentum-conserving delta-function
$$
\D \equiv  i (2 \pi )^{8} \d^{8} (\sum p_i).
$$
All momenta are incoming and the arrows indicate the flow of
quantum $\hat{\bZ}_N$ charge.

The asymptotics of $I(z,\bar z)$ as $z \to 0$ or $z \to 1$ imply
that $V_4$ does not have the usual pole structure in the  $s$ and
$t$ channels. This is a consequence of the noncompactness of the
cone. In the $s$ and $t$ channels, the exchanged states are
untwisted states which have momentum along the cone directions
$(X, \bar X)$ and a continuous mass spectrum from
eight-dimensional point of view. These states can contribute
because there is no translation invariance in those directions and
hence momentum is not conserved. As a result, the poles are
replaced by softer logarithmic divergences. The leading
$s$-channel behavior comes from integrating near $z=0$,
\be V_4^s \approx -{{g'_c}^4 C \over 2\pi \a'}\D (\a' p_1 \cdot
p_3/2)^2 \int_0 d^2 z {|z|^{-{\a' s \over 2}- 2} \over \log {|z|
\over \d}}. \label{schannel} \ee
Integrating near $z=1$ gives the leading $t$-channel contribution
\be V_4^t \approx -{{g'_c}^4 C \over 2\pi \a'} \D (\a' p_1 \cdot
p_3/2)^2 \int_0 d^2 z {|z|^{-{\a' t \over 2}- 2} \over \log {|z|
\over \d}}. \label{tchannel} \ee
In the $u$ channel, which comes from the $z \to \infty$ region of
the integral, the exchanged states are localized twisted sector
states and there one gets a sum over pole terms. There are  no
massless (or nearly massless) poles in the $u$-channel; the first
contribution comes from a massive exchange
\be V_4^u \approx -{2{g'_c}^4 C  \over \pi  \a'}\tan ({\p k \over
N}) \D (\a' p_1 \cdot p_3/2)^2 {\G^4 ({k \over N}) \over \G^2 ({2k
\over N}-1)} {1\over \a' u / 2 + 2 (2 - 3 k/N)} . \label{uchannel}
\ee

\subsection{Cubic Couplings to Other Twisted States}\label{Other}

We now show that, at least in the  quartic approximation to the
tachyon potential, a constant vev for one of the nearly-marginal
tachyons does not generate a tadpole for any of the other
nearly-massless fields in the twisted sectors. It is therefore
consistent to neglect these states in the analysis to quartic
order.

The couplings between twisted sector states are severely
restricted by quantum $\hat{\bZ}_N$ symmetry and $H$-charge
conservation. Let us start with the three-point couplings.
Possible tadpoles come from couplings $\langle \bar T_k^{(-1,-1)}
\bar T_k^{(-1,-1)} \F^{(0,0)} \rangle$ where the superscript
denotes the picture. Quantum symmetry and $H$-charge conservation
imply that the zero-picture state $\F^{(0,0)}$ has to be
proportional to $\s_{2 k -N} e^{i 2 k/N (H-\tilde H)}$. The lowest
state with these quantum numbers has vertex operator
$$
\F^{(0,0)} = e^{ip_\m X^\m} \s_{2 k -N} e^{i 2 k/N (H-\tilde H)} c \tilde c
$$
which has mass-squared $m^2 = 4/\a' ( -2 + 3 k/N)$ which is  of
order of the string scale if $k/N$ is close to one  for example
when $k =N-2$. This is consistent with what we find in equation
(\ref{uchannel}). It shows that the lowest lying exchanged state
in the $u$-channel has precisely this mass, and there are no poles
from exchanging tachyonic or nearly massless fields.

There are four-point couplings of the form $\langle \bar
T_k^{(-1,-1)} T_k^{(0,0)} \bar T_k^{(-1,-1)} \F^{(0,0)} \rangle$
which could also source other nearly-massless tachyons. Again
using quantum symmetry and $H$-charge conservation one sees that
$\F^{(0,0)}$ is either equal to $T_k^{(0,0)}$ or proportional to
$\s_k e^{i(k/N+1) (H- \tilde H)}$. The lowest mass state with the
latter quantum numbers has vertex operator
$$ \F^{(0,0)} =  e^{ip_\m X^\m}
\s_k e^{i(k/N+1) (H- \tilde H)} c \tilde c
$$
with $m^2 = 2/\a' ( -1 + 3 k/N)$. This is again a massive state
with string scale mass for $k/N \approx 1$.

It seems possible to generalize  this argument for at least a
large class of higher point functions but we restrict ourselves
only up to quartic order.

\subsection{Cubic Coupling to Untwisted Massless Fields} \label{cubic}

We now calculate the cubic vertex for two tachyons and one
massless field from the untwisted sector. The vertex operator for
a massless state with polarization tensor $e_{MN}$ in the zero
picture is
\be H^{(0,0)}(z, \bar z,p, e) \equiv - g_c  e_{MN} {2 \over \a '}
\left[(i \pa X^M + {\a ' \over 2} p\cdot \psi \psi^M) (i \bar \pa
X^N + {\a' \over 2} p\cdot \tilde \psi \tilde \psi^N) c\tilde c
e^{i p \cdot X} (z, \bar z)\right]_N. \nonu \label{massless} \ee
Gravitons are described by a symmetric, traceless polarization
tensor and B-field fluctuations correspond to an antisymmetric
polarization tensor. The dilaton vertex operator requires a bit
more care. In the $(-1)$ picture, it is given by
\cite{Polchinski:1988tu}
$$
{1 \over \sqrt{8}}\left[ ( \psi \cdot \tilde \psi e^{- \f- \tilde
\f } - \pa \x \tilde \h e^{-2 \f} - \pa \tilde \x \h e^{-2 \tilde
\f})c \tilde c e^{i p \cdot X} \right]_N
$$
Applying the picture changing operator to this, we find that the
zero-picture dilaton vertex operator is given by (\ref{massless})
with $e_{MN} = {1 \over \sqrt{8}} \h_{MN}$ plus terms with either
$\f$-charge different from zero or ghost number different from
one. Such terms don't contribute to the three-point amplitude.

Using the three-point functions from (\ref{ccoeffs}) with
$\e_1=\e_2=-1$ one finds the cubic coupling
\bea V_3(p_1,p_2;p_3,e) &=& \langle
T_k^{(-1,-1)} (z_\infty, \bar z_\infty, p_1) \bar T_k^{(-1,-1)}
 (1, 1, p_2) H^{(0,0)}(0,0,p_3, e)\rangle\nonumber\\
&=& - {\a' {g'_c}^2  g_c C  \over 2} \d^{- \a' p_3 \bar
p_3\over 2} \D
\nonumber\\
&&\Big(e_{\m \n} p_2^\m p_2^\n - e_{\m  X}  p_2^\m \bar p_3 -
e_{\bar X \m}  p_2^\m p_3 + e_{\bar X X} p_3 \bar p_3 \Big)
\label{3point} \eea Note that this expression is not symmetric in
the polarization indices; this means that there is a coupling to
the B-field as well as to the graviton and dilaton.
 One way to motivate the choice $\e_1=\e_2=-1$
in (\ref{ccoeffs}) is that it is the only one that leads to a
BRST-invariant amplitude; indeed one easily checks that
(\ref{3point}) is invariant under \be e_{MN} \to e_{MN} + p_{3M}
a_N + p_{3N} b_M \label{gaugetransf} \ee upon using $p_1^2 = p_2^2
= -m^2,\ p_3^2=0$. In (\ref{3point}), the graviton is in the
transverse-traceless gauge. In order to compute the graviton
exchange diagram we would like to know the correct vertex to use
in the harmonic gauge $p^M e_{(MN)} - \half p_N e^M_M =0$ for the
graviton. In this gauge there is more residual gauge invariance
and one would typically expect to have to add terms proportional
to $p^M e_{(MN)}$  and $e^M_M $ until the vertex is invariant
under the larger set of gauge transformations. In our case however
we saw that (\ref{3point}) is  already invariant under
(\ref{gaugetransf}) without imposing $a\cdot p_3 = b\cdot p_3=0$.
Hence (\ref{3point}) is the correct vertex to use in the harmonic
gauge. Another proof of the validity of (\ref{3point}) in the
harmonic gauge will be given in \cite{paper2}.

The final form of the cubic coupling (\ref{3point}) shows that the
tachyon has a Gaussian form factor in its coupling to the massless
untwisted fields. At large $N$, the width of the Gaussian in
position space scales as $\sqrt{N}$. The coupling is thus not
point-like but spread over a very large radius of order $\sqrt{N}$
times the string length. Because the opening angle of the cone is
also getting smaller as $1/N$, the total area over which this
interaction takes place is still of order one in string units.

\section{Unitarity and Massless Exchanges}
\label{Factorization}

\FIGURE{\epsfig{file=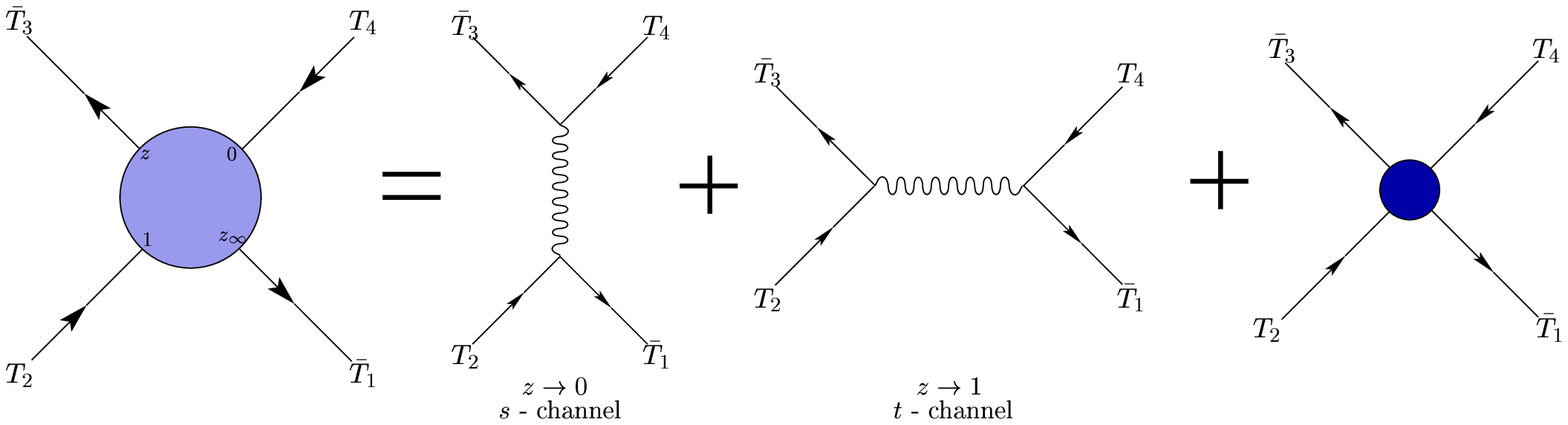,width=15cm}\caption{Factorization
and the quartic contact interaction.}}
In this section,  we first determine the over-all normalization
from unitarity in $\S{\ref{Determine}}$, and then compute the
massless exchange diagrams in $\S{\ref{Massless}}$. By subtracting
these exchanges from the four tachyon scattering amplitude we can
extract a possible quartic contact term. The Feynman diagram is
shown in figure 2. We find that in the $u$-channel only massive
particles of  string-scale mass are exchanged consistent with
(\ref{uchannel}). Hence we need to subtract only the $s$ and $t$
channel exchanges. As explained in $\S{\ref{Strategy}}$, the
quartic contact term on the right hand side of figure 2 is the
effective quartic term at low energy  and includes the exchanges
of particles of string scale mass.

\subsection{Determination of Normalization
Constants}\label{Determine}

So far we have introduced a number of  undetermined normalization
constants:

$C$: overall normalization of the path integral on the sphere.

$g_c$: normalization of the graviton vertex operator or the
`closed string coupling'.

$g_c'$: normalization of the tachyon vertex operator.\\
Note that the constant $A$ that was introduced  for the $X
\bar{X}$ CFT in (\ref{sigmanorm}) is absorbed in $C$ along with
other functional determinants and the constant $B$ is already
determined in (\ref{B}).

Unitarity of the S-matrix allows us to express all constants in
terms of $\a '$ and $g_c$. The latter is in turn proportional to
the gravitational constant $\k$. We now work out these relations
keeping track of possible factors of $N$.

The constant $C$ can be expressed in terms of $\a '$ and $g_c$ by
factoring the four-tachyon amplitude on graviton exchange. From
(\ref{3point}) we calculate the contribution to the 4-point
function coming from  the exchange of longitudinal gravitons,
\bea
V_4^{exch} &=&- i  \int { d^8 p^\m \over (2 \p)^{8}}\int_\bC
{dp d \bar p \over (2 \p)^{2}}
{V^3_{\m\n}(p_1,p_2;p)V^{3\m\n}(p_3,p_4;p)
\over p^\m p_\m + 2 p \bar p}\label{factor}\\
&&= -{g'_c}^4 g_c^2 C^2 { 1 \over 16  \p^2} \D
 (\a' p_1 \cdot p_3/2)^2
\int_0 d^2 z  {|z|^{-{\a' s \over 2}- 2} \over \log {|z| \over
\d}}\, . \label{factor2} \eea
In writing the momentum space propagator in the first line, we
have assumed a specific normalization for the spacetime field
created by the vertex operator with normalization $g_c$. The
normalization we used is appropriate for a canonically normalized
field on the covering space $\bM^8 \times \bC$ which is periodic
under $x \to e^{2 \p i/N} x$. The details are explained in
appendix \ref{Feynman}. Comparing (\ref{factor2}) with
(\ref{schannel}) we find the overall normalization
\be C = {8  \p \over \a' g_c^2} \label{C} \, . \ee
This is the familiar flat-space value as it should be by
construction.

The normalization of the tachyon vertex operator $g'_c$ can be
determined in terms of $g_c$ by factoring the 2 graviton-2 tachyon
amplitude on the pole coming from tachyon exchange:
 \be W_4 \equiv \int d^2 z \langle
T_k^{(0,0)} (z_\infty, \bar z_\infty, p_1)H^{(-1,-1)}(1, 1,p_2,
e_2) \bar T_k^{(0,0)} (z,\bar z , p_3)H^{(-1,-1)}(0, 0,p_4, e_4)
\rangle\label{W} \ee
 For simplicity
we can take the graviton to have polarization along the
longitudinal $X^\m$ directions. For $z \to 0$, there is a pole at
$s = -2/\a '(1 - k/N) $ coming from tachyon exchange. We can find
the coefficient at the pole from the OPE of $\bar T_k^{(0,0)}$
with $H^{(-1,-1)}$. From the 3-point function (\ref{ccoeffs}) we
know that
$$
\s_{N-k}(z) [e^{i(px + \bar p \bar x})]_N(0) \sim |z|^{-\a ' p
\bar p}
 \d^{-\a 'p \bar p/2} \s_{N-k}
$$
Using this we find the OPE
$$
\bar T_k^{(0,0)} (z,\bar z  , p_3)H^{(-1,-1)}(0, 0,p_4, e_4)$$
$$\sim {\a ' \over 2} g_c g_c' e_{4\m\n}p_3^\m p_3^\n |z|^{-s-m^2-2}
 \d^{-\a ' p_4 \bar p_4}  e^{i(p_3+p_4)_\m X^\m} \s_{N-k}
 e^{-i/N(H-\tilde H)}
 c \tilde c e^{-\f-\tilde \f}(0)
$$
Substituting in (\ref{W}) and integrating around $z=0$ gives the
pole term
$$
W_4 \sim { - 2 \p \a' g_c^2 {g'_c}^2 C \d^{- \a ' /2( p_3\bar p_3
+ p_4 \bar p_4)} ( e_{2 \m \n} p_1^\m p_1^\n ) (e_{4 \r \s} p_3^\r
p_3^\s ) \over  s + 2/\a ' (1 - k/N)} \D
$$
Comparing with (\ref{3point}) we get
$$
{g'_c}^2 C = {8 \p \over \a'}
$$
and hence \be g_c' = g_c. \label{gc'} \ee
An extra check of these relations is provided in appendix
\ref{Check} where we compute the first massive exchange in the
$u$-channel and find agreement with (\ref{uchannel}).

We have not yet determined the proportionality constant between
the vertex operator normalization $g_c$ and the gravitational
coupling $\k$. To compare with the prediction (\ref{conjecture})
we should use $\k$ which is the cubic coupling for gravitons
canonically normalized on $\bM^8 \times \bC$. Hence, it is related
to the closed string coupling as usual by
\be \k  = 2 \p g_c. \label{kappa} \ee
%

\subsection{Massless Exchange Diagrams}\label{Massless}

Having obtained the cubic vertex for two tachyons and a massless
field in (\ref{3point}), we can calculate the contribution of
massless exchange diagrams to the 4-tachyon amplitude. These
diagrams will contain integrals over the momentum along the cone;
they will be  of the form \bea I_n (s) &\equiv& \int_{\bf C} {dp d
\bar p \over (2\p)^2}
 {(p\bar p)^{n-1} \d^{-\a ' p\bar p} \over - s +
2 p \bar p} \nonu &= & - {(-\a ')^{1-n} (n-1)!\over 16 \pi^2}
\int_{D_1} d^2 z
 {|z|^{-{\a' s \over 2}- 2} \over (\log {|z| \over \d})^n}.
\eea The second form is useful for comparing with the string
amplitude (\ref{4point}). The domain $D_1$ is  the unit disc.

{\bf Dilaton exchange}

The vertex is
$$
V_3^{dil}(p_1,p_2;p_3)= -{ \k \over \sqrt{2}} ( p_3 \bar p_3 -
m^2)
 \d^{- \a' p_3 \bar p_3 /2}
$$
and the dilaton propagator is given by:
$$
- {i \over p^M p_M } .
$$
Hence we find the exchange amplitude
$$
V_4^{exch,\ dil} = {4 \k^2  \over 8} \D (m^4 I_1 - 2 m^2 I_2 +
I_3)
$$

{\bf B-field exchange}

The vertex is
$$
V_3^{B}(p_1,p_2;p_3,e)=  2 \k  \D \left( e_{[\m  x]} p_2^\m \bar
p_3 + e_{[\bar x \m]}  p_2^\m p_3 + e_{[x \bar x]} p_3 \bar p_3
\right) \d^{- \a' p_3 \bar p_3 /2}
$$
The propagator is, in the Feynman  gauge $p^M e_{[MN]} = 0$,
$$
- {i \over p^M p_M } \left( \half \h_{MR} \h_{NS}- \half \h_{MS}
\h_{NR}\right).
$$
This gives the exchange amplitude
$$
V_4^{exch,\ B} = - 4 \k^2  \D ( p_2\cdot p_4 I_2 + \half I_3).
$$

{\bf Graviton exchange}

The vertex is given by
\bea V_3^{grav}(p_1,p_2;p_3,e)&=& -2 \k  \D \Big(e_{(\m \n)}
p_2^\m p_2^\n - e_{(\m  x)}  p_2^\m \bar p_3 - e_{(\bar x \m)}
p_2^\m p_3\nonu &&+ \half  e_{(\bar x x)} p_3 \bar p_3 \Big)
 \d^{- \a' p_3 \bar p_3 /2}.
\eea
The propagator in the harmonic gauge $p^M e_{(MN)} - \half p_N
e^M_M =0$ reads
$$
 - {i \over p^M p_M } \left( \half \h_{MR} \h_{NS}+
\half \h_{MS} \h_{NR} - {1 \over 8}  \h_{MN} \h_{RS} \right).
$$
This gives the exchange amplitude
\bea V_4^{exch,\ grav} &=&  4 \k^2  \D \Big( (( p_2\cdot p_4)^2 -
{m^4 \over 8}) I_1 \nonu && + {m^2 \over 4} I_2 + p_2 \cdot p_4
I_2 + \half I_3  - {1 \over 8} I_3 \Big). \eea

\subsection{Subtractions and quartic term for the tachyon}

Summing these contributions, we see that many terms cancel and we
are left with
\be V_4^{exch,\ total} = 4 \k^2  \D (p_1 \cdot p_3)^2 I_1 (s).
\label{exch} \ee
A similar term comes from the massless $t$-channel exchanges.
These contributions yield precisely  the asymptotics of the string
amplitude (\ref{schannel}) without extra terms finite at zero
momentum. Such terms, if present, would have contributed to the
quartic contact term for the tachyon. In fact, for the open string
system studied in \cite{Gava:1997jt}, the massless subtractions do
yield such extra terms and, in that case, they give the leading
contribution to the quartic tachyon potential.

The quartic tachyon coupling is given by  integral (\ref{4point})
with the massless exchanges subtracted. The coefficient in front
of the integral reads
$$ {g_c'^4 C \a ' \over 4 \p^2} \sin{\p k\over N} (p_1 \cdot p_3)^2
= {\kappa^2 \over 2 \p^3}  \sin{\p k\over N} (p_1 \cdot p_3)^2$$
which is of order $1/N^3$. We shall now show that the remaining
integral is of order one. It is given by
$$
J= \int_{\bf C} d^2 z | 1- z|^{-{\a 't \over 2} -2} |z|^{-{\a' s
\over 2} - 2} I(z, \bar z)^{-1} + {\p \over 2 \sin (\p k /N)
}\int_{D_1} {|z|^{-{\a' s \over 2} - 2} +|z|^{-{\a' t \over 2} -
2} \over \ln {|z| \over \d}}.
$$
The second term comes from subtracting the massless exchanges
(\ref{exch}). In evaluating the first term in $J$ for large $N$,
one should be careful because the hypergeometric function $I$ does
not converge uniformly. The function $I$ approaches the value 2
everywhere except in the points $z=0,1$ (see (\ref{asymptotics})).
In evaluating the integral numerically, one finds numerical
convergence problems in the regions around $z=0,1$. A similar
situation was encountered in \cite{Klebanov:2003my}. We therefore
split the integral in three parts: we cut out two small discs of
radius $\e$ around $z=0,1$ where we approximate the integrand by
its asymptotics (\ref{asymptotics}) which we can integrate
analytically. The integral over the remainder of the complex plane
will be easy to evaluate numerically using Mathematica. After
summing the three integrals and subtracting the massless exchanges
we take $\e$ to zero.

Let us start with the integral near zero with the $s$-channel
exchanges subtracted. The result is an integral over the unit disc
with a small disc around $z=0$ removed: \bea J_{z=0} &=& {2 \p^2
\over \sin (\p k /N)} \int_\e^1 d r {r^{-{\a' s \over 2} - 1}
\over \ln {|z| \over \d}}\nonu &=& {2 \p^2 \over \sin (\p k
/N)}\d^{-{\a' s \over 2}}\left(E_1 \left( {\a's \over 2} \ln {\e
\over \d} \right) - E_1 \left( -{\a's \over 2} \ln \d
\right)\right)\nonu &\approx& 2 \p \ln \e  + \co(1/N^2) \eea Here,
$E_1$ is the exponential integral $E_1(x) = \int_x^\infty dt
e^{-t}/t$ and, in the last line, we have displayed the leading
term at large $N$. The integral around $z=1$ with the $t$-channel
exchanges subtracted has the same leading behavior. So the leading
term for $J$ is
$$
J \approx 4 \p \ln \e + \half \int_{\bf C \backslash {\rm discs}}|
1- z|^{ -2} |z|^{ - 2}
$$
\FIGURE{
\begin{picture}(280,160)
\put(19,156){$J$} \put(255,122){$1/ \e$}
\put(0,0){\epsfig{file=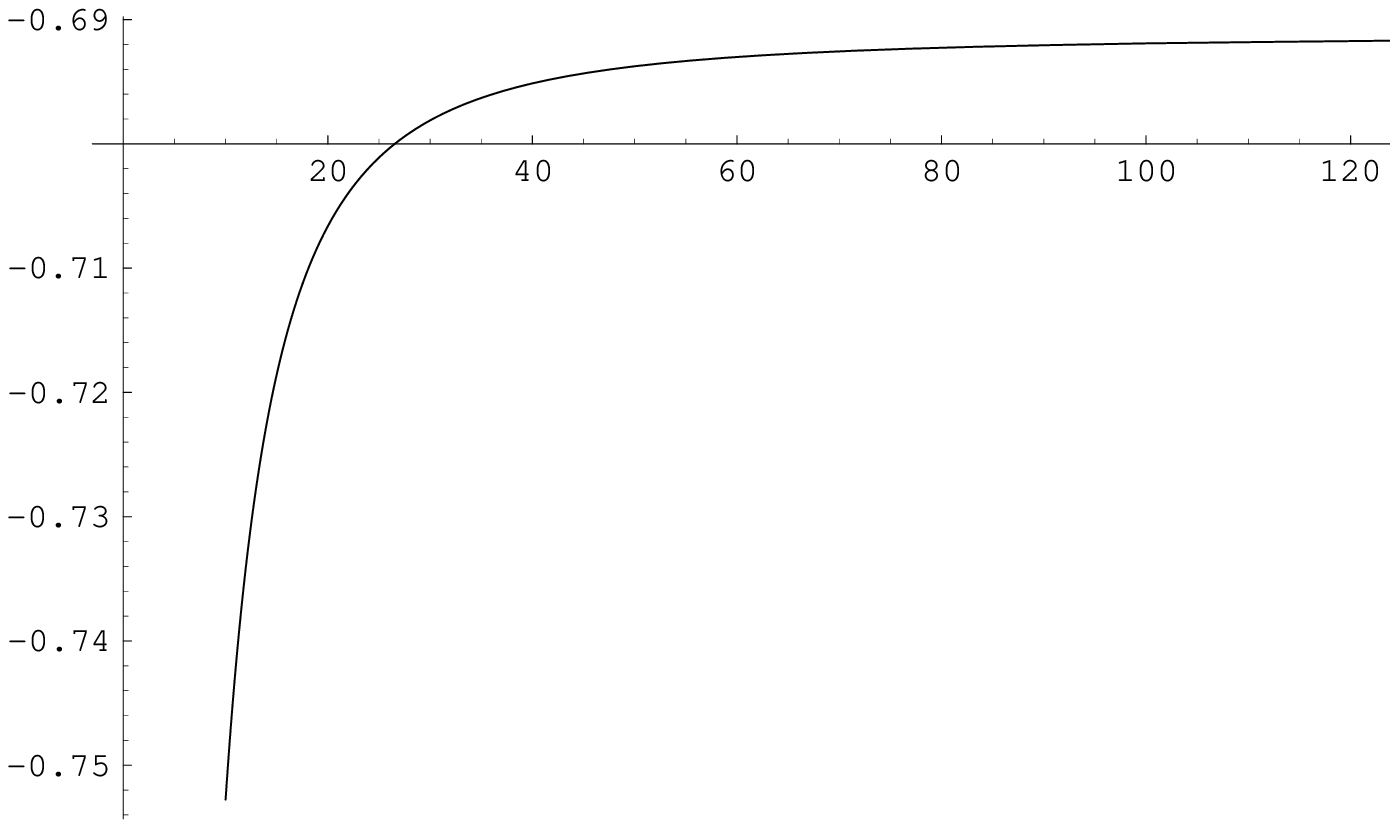, width=250pt}}
\end{picture}
\caption{The integral $J$ for $s,t=0$ and $N=\infty$ as a function
of $1/ \e$.}\label{plot}}
The integral runs over the complex plane with small discs around
$z=0,1$ removed. We have used that, in this integration region,
the function $I$ uniformly approaches the value of $2$ at large
$N$. We have also taken $s=t=0$. The result of the numerical
integration is plotted as a function of $1/\e$ in figure
\ref{plot}. As $\e$ goes to zero, $J$ converges to
$$
J \approx -0.691.
$$

\section{Conclusions and Comments}\label{Conclusions}

We have seen that the system of localized tachyons in \cn
backgrounds provides a tractable system to study aspects of
off-shell closed string theory. Condensation of these tachyons
connects all \cn backgrounds to each other and to flat space.
There is a well-defined conjecture for the height of the potential
that is rather analogous to the open-string case. Moreover,
considerable computational control is possible using orbifold CFT
techniques.

We have analyzed the system in the large-$N$ approximation where
it is possible to read off the off-shell action from the S-matrix.
We have been able to compute  all three-point correlation
functions required to describe the interaction of the tachyons
with massless untwisted fields. Motivated by a simple model of the
tachyon potential we compute the quartic contact term.  If the
quartic contact term is of order one, then the minimum can occur
very close to the origin and higher point interactions can be
consistently ignored. Our calculation however yields a  quartic
term that is much too small and goes instead  as $1/N^3$. This
implies that our simple model is not valid for describing the
potential.

There are a number of possible ways to get around this problem.
One possibility is that by going  beyond quartic order one can
find the new minimum of the potential of the desired depth. It is
not clear however how one can obtain a very shallow minimum if
higher order terms are important. Another more likely possibility
is that the direction that we have chosen in the field space of
tachyons is not the correct one for finding the minimum. Our
choice of this specific tachyon was guided by the analysis of
\cite{Harvey:2001wm, Vafa:2001ra} which indicates that to go from
\cn orbifold to the $\bC/\bZ_k$ orbifold by RG-flow, one needs to
turn on a specific relevant operator of definite charge $k$ under
the quantum symmetry. Our tachyon corresponds precisely to this
relevant operator near the conformal point. This assumption is
further supported by our finding in $\S{\ref{Other}}$ that turning
on the tachyon of charge $k$ does not source other tachyons and
thus  is a consistent approximation.

It is possible however that other excited tachyons are also
involved in this process. There are a number excited states in the
twisted sectors that are nearly massless or massless. The analysis
of \cite{Vafa:2001ra} is not sensitive to excited tachyons because
it deals with only the chiral primaries. In particular, the
analysis of \cite{Adams:2001sv} shows that to go from \cn to
$\bC/\bZ_k$, the operator that is turned on in the quiver theory
does not have definite charge under the quantum symmetry. It would
be important to understand how these two pictures -- the D-probe
analysis and the RG-flows -- are consistent with each other. This
might help in identifying all the fields that are involved in the
condensation \cite{paper2}.

A computation of the height of the potential for the \cn tachyons
was attempted earlier in \cite{Sarkar:2003dc} where a large-$N$
approximation was made inside the integral over the worldsheet
coordinate $z$.  As we have seen, this approximation is, however,
not uniform over the region of integration and breaks down near
$z=0, 1$. As a result one obtains poles for massless exchanges
instead of the correct softer logarithmic behavior that we have
found. Moreover, the subtractions made in \cite{Sarkar:2003dc}
were based on a postulated effective action  which differs
substantially from the actual cubic interactions that result from
our computations.

\acknowledgments

We would like to thank A.~Adams, M.~Headrick, S. Minwalla, E.
Martinec, Y. Okawa, S.~Shenker, S.-J. Sin, T. Suyama, C. Vafa and
B. Zwiebach  for useful discussions. The work of A. D. was
supported in part by Stanford Institute of Theoretical Physics,
the David and Lucile Packard Foundation Fellowship for Science and
Engineering, and by the Department of energy under contract number
DE-AC03-76SF00515. J. R.~thanks the Department of Theoretical
Physics at the Tata Institute of Fundamental Research where this
work was initiated and the hospitality of Harvard University and
the SLAC theory group, where part of this work was completed.

\begin{appendix}

\section{Feynman Rules on the Cone}\label{Feynman}
In order to compare string amplitudes with the momentum space
Feynman rules of a field theory on \cn, it is convenient to mimic
the construction of $\S{\ref{CFT}}$ and work with fields $\f$
defined on {\bf C} and with periodicity \be \f( \eta x, \bar \eta
\bar x) = \f(x, \bar x) \label{per} \ee
 with $\eta = e^{2 \p i /N}$.
Suitable basis states on the cone are \be \Psi_{p, \bar p} (x,
\bar x) = 1/N \sum_{k=0}^{N-1} \exp i(\eta^k p x + \bar \eta^k
\bar p \bar x) \label{normalbasis} \ee These are normalized just
like the basis states of the CFT (\ref{norm}):
$$
\int_{\bf C} dx d \bar x  \Psi^*_{p', \bar p'} \Psi_{p, \bar p}
=(2\p )^2 \d_N(\vp,\vp ').
$$
Momentum space Feynman rules are obtained by expanding fields in
this basis:
$$
\f(x, \bar x) = \int_\bC {dp d\bar p \over (2 \p)^2}\tilde \f(p, \bar p)  \Psi_{p, \bar
  p}
(x,\bar x)
$$
In writing the factorization (\ref{factor}) we have assumed that
the string vertex operators create spacetime fields with
propagator $p^\m p_\m + 2 p \bar p$. Hence we should compare the
string amplitudes with Feynman rules for spacetime fields $\f$
with canonically normalized kinetic term \bea S_{kin} &=& \int d^8
x \int_{\bf C} d x d \bar x \f (\pa^\m \pa_\m +2 \pa \bar \pa) \f.
\label{cannorm} \nonu &=& \int d^8 p^\m \int_\bC {dp d \bar p\over
(2 \p)^2} \tilde \f(p^\m p_\m + 2 p \bar p) \tilde \f. \eea

\section{Consistency Check: Massive Exchange}\label{Check}

We saw that there is a nonvanishing coupling $\langle \bar
T_k^{(-1,-1)} \bar T_k^{(-1,-1)} \F^{(0,0)} \rangle$ with
$\F^{(0,0)}$ a state with mass $m^2 = 4/\a' ( -2 + 3 k/N)$. We
shall now show that its exchange diagram is in agreement with the
string result (\ref{uchannel}).

First, we need the three-twist correlator (for $k/N > \half$)
$$
C_{-,-,++} \equiv \langle \s'_{N-k}(\infty) \s_{N-k}(1)
\s_{2N-2k}(0) \rangle
$$
This can be obtained from the $z\to \infty$ factorization limit
of (\ref{Z}) which has intermediate states in the twisted
sector. Using the asymptotics (\ref{asymptotics}), the result is
\be |C_{-,-,++}| = A
\sqrt{2 \over \a'}{\sqrt{|\tan \p k/N|}\over 2 \p}
 {\G^2(k/N) \over |\G( 2k/N-1)|}.
\label{3twist} \ee This agrees with the finite volume result
(formula (4.47) in \cite{Dixon:1987qv}) upon taking $\a'=2$ and
the momentum space volume $V_\L = 1/4 \p^2$.

In the $(-1,-1)$ picture, the
vertex operator for the massive state is
$$
\F^{(-1,-1)} = {N \over 2 k -N}  \a_{- {2k-N \over N}} \bar{\tilde \a}_{- {{2k-N} \over N}}
\cdot \s_{2 k -N} e^{i{2k-N \over N}(H-\tilde H)} e^{i p^\m X_\m}c \tilde c e^{-\f - \tilde \f}
$$
One can check that this is indeed a physical state. In the $(0,0)$ picture,
the relevant part of the vertex operator is
$$
\F^{(0,0)} = {2 k - N \over N}\s_{2 k -N} e^{i{2k\over N}(H-\tilde H)} e^{i p^\m X_\m}c \tilde c
+ \ldots
$$
Using the bosonic amplitude (\ref{3twist}), we find for the three-point amplitude
\bea
M_3 &\equiv& \langle  \bar T_k^{(-1,-1)}\; '(\infty,p_1) \bar T_k^{(-1,-1)}(1,p_2) \cv^{(0,0)}
(0,p_3) \rangle\nonu
&=& g_c'^3 C \sqrt{2 \over \a'} {2 k-N\over N} \D
{\sqrt{|\tan \p k/N|}\over 2 \p}
 {\G^2(k/N) \over |\G(2k/N-1)|}
\eea
Note that one cannot decide from the on-shell 3-point function
whether the coupling is derivative or nonderivative. The exchange
diagram reads \be A_4^{massive \ exch} = - g_c'^6 C^2 ({2 k-N\over
N})^2 {|\tan \p k/N|\over 4 \p^2} {\G^4(k/N) \over \G^2(
2k/N-1)}\D {2/\a ' \over u- 4/ \a' ( -2 + 3k/N)}
\label{massiveexch} \ee This agrees with the string amplitude
(\ref{uchannel}) at the $u$-channel pole; we also see from
(\ref{uchannel}) that the three-point coupling is in fact a
 derivative one.
\end{appendix}

\bibliography{ref}
\bibliographystyle{utphys}

\end{document}